\documentclass{article}
\usepackage{times}
\usepackage{graphicx} % more modern
\usepackage{subfigure} 
\usepackage[draft]{minted}

\makeatletter
\def\PYG@reset{\let\PYG@it=\relax \let\PYG@bf=\relax%
    \let\PYG@ul=\relax \let\PYG@tc=\relax%
    \let\PYG@bc=\relax \let\PYG@ff=\relax}
\def\PYG@tok#1{\csname PYG@tok@#1\endcsname}
\def\PYG@toks#1+{\ifx\relax#1\empty\else%
    \PYG@tok{#1}\expandafter\PYG@toks\fi}
\def\PYG@do#1{\PYG@bc{\PYG@tc{\PYG@ul{%
    \PYG@it{\PYG@bf{\PYG@ff{#1}}}}}}}
\def\PYG#1#2{\PYG@reset\PYG@toks#1+\relax+\PYG@do{#2}}

\expandafter\def\csname PYG@tok@gd\endcsname{\def\PYG@tc##1{\textcolor[rgb]{0.63,0.00,0.00}{##1}}}
\expandafter\def\csname PYG@tok@gu\endcsname{\let\PYG@bf=\textbf\def\PYG@tc##1{\textcolor[rgb]{0.50,0.00,0.50}{##1}}}
\expandafter\def\csname PYG@tok@gt\endcsname{\def\PYG@tc##1{\textcolor[rgb]{0.00,0.27,0.87}{##1}}}
\expandafter\def\csname PYG@tok@gs\endcsname{\let\PYG@bf=\textbf}
\expandafter\def\csname PYG@tok@gr\endcsname{\def\PYG@tc##1{\textcolor[rgb]{1.00,0.00,0.00}{##1}}}
\expandafter\def\csname PYG@tok@cm\endcsname{\let\PYG@it=\textit\def\PYG@tc##1{\textcolor[rgb]{0.25,0.50,0.50}{##1}}}
\expandafter\def\csname PYG@tok@vg\endcsname{\def\PYG@tc##1{\textcolor[rgb]{0.10,0.09,0.49}{##1}}}
\expandafter\def\csname PYG@tok@vi\endcsname{\def\PYG@tc##1{\textcolor[rgb]{0.10,0.09,0.49}{##1}}}
\expandafter\def\csname PYG@tok@vm\endcsname{\def\PYG@tc##1{\textcolor[rgb]{0.10,0.09,0.49}{##1}}}
\expandafter\def\csname PYG@tok@mh\endcsname{\def\PYG@tc##1{\textcolor[rgb]{0.40,0.40,0.40}{##1}}}
\expandafter\def\csname PYG@tok@cs\endcsname{\let\PYG@it=\textit\def\PYG@tc##1{\textcolor[rgb]{0.25,0.50,0.50}{##1}}}
\expandafter\def\csname PYG@tok@ge\endcsname{\let\PYG@it=\textit}
\expandafter\def\csname PYG@tok@vc\endcsname{\def\PYG@tc##1{\textcolor[rgb]{0.10,0.09,0.49}{##1}}}
\expandafter\def\csname PYG@tok@il\endcsname{\def\PYG@tc##1{\textcolor[rgb]{0.40,0.40,0.40}{##1}}}
\expandafter\def\csname PYG@tok@go\endcsname{\def\PYG@tc##1{\textcolor[rgb]{0.53,0.53,0.53}{##1}}}
\expandafter\def\csname PYG@tok@cp\endcsname{\def\PYG@tc##1{\textcolor[rgb]{0.74,0.48,0.00}{##1}}}
\expandafter\def\csname PYG@tok@gi\endcsname{\def\PYG@tc##1{\textcolor[rgb]{0.00,0.63,0.00}{##1}}}
\expandafter\def\csname PYG@tok@gh\endcsname{\let\PYG@bf=\textbf\def\PYG@tc##1{\textcolor[rgb]{0.00,0.00,0.50}{##1}}}
\expandafter\def\csname PYG@tok@ni\endcsname{\let\PYG@bf=\textbf\def\PYG@tc##1{\textcolor[rgb]{0.60,0.60,0.60}{##1}}}
\expandafter\def\csname PYG@tok@nl\endcsname{\def\PYG@tc##1{\textcolor[rgb]{0.63,0.63,0.00}{##1}}}
\expandafter\def\csname PYG@tok@nn\endcsname{\let\PYG@bf=\textbf\def\PYG@tc##1{\textcolor[rgb]{0.00,0.00,1.00}{##1}}}
\expandafter\def\csname PYG@tok@no\endcsname{\def\PYG@tc##1{\textcolor[rgb]{0.53,0.00,0.00}{##1}}}
\expandafter\def\csname PYG@tok@na\endcsname{\def\PYG@tc##1{\textcolor[rgb]{0.49,0.56,0.16}{##1}}}
\expandafter\def\csname PYG@tok@nb\endcsname{\def\PYG@tc##1{\textcolor[rgb]{0.00,0.50,0.00}{##1}}}
\expandafter\def\csname PYG@tok@nc\endcsname{\let\PYG@bf=\textbf\def\PYG@tc##1{\textcolor[rgb]{0.00,0.00,1.00}{##1}}}
\expandafter\def\csname PYG@tok@nd\endcsname{\def\PYG@tc##1{\textcolor[rgb]{0.67,0.13,1.00}{##1}}}
\expandafter\def\csname PYG@tok@ne\endcsname{\let\PYG@bf=\textbf\def\PYG@tc##1{\textcolor[rgb]{0.82,0.25,0.23}{##1}}}
\expandafter\def\csname PYG@tok@nf\endcsname{\def\PYG@tc##1{\textcolor[rgb]{0.00,0.00,1.00}{##1}}}
\expandafter\def\csname PYG@tok@si\endcsname{\let\PYG@bf=\textbf\def\PYG@tc##1{\textcolor[rgb]{0.73,0.40,0.53}{##1}}}
\expandafter\def\csname PYG@tok@s2\endcsname{\def\PYG@tc##1{\textcolor[rgb]{0.73,0.13,0.13}{##1}}}
\expandafter\def\csname PYG@tok@nt\endcsname{\let\PYG@bf=\textbf\def\PYG@tc##1{\textcolor[rgb]{0.00,0.50,0.00}{##1}}}
\expandafter\def\csname PYG@tok@nv\endcsname{\def\PYG@tc##1{\textcolor[rgb]{0.10,0.09,0.49}{##1}}}
\expandafter\def\csname PYG@tok@s1\endcsname{\def\PYG@tc##1{\textcolor[rgb]{0.73,0.13,0.13}{##1}}}
\expandafter\def\csname PYG@tok@dl\endcsname{\def\PYG@tc##1{\textcolor[rgb]{0.73,0.13,0.13}{##1}}}
\expandafter\def\csname PYG@tok@ch\endcsname{\let\PYG@it=\textit\def\PYG@tc##1{\textcolor[rgb]{0.25,0.50,0.50}{##1}}}
\expandafter\def\csname PYG@tok@m\endcsname{\def\PYG@tc##1{\textcolor[rgb]{0.40,0.40,0.40}{##1}}}
\expandafter\def\csname PYG@tok@gp\endcsname{\let\PYG@bf=\textbf\def\PYG@tc##1{\textcolor[rgb]{0.00,0.00,0.50}{##1}}}
\expandafter\def\csname PYG@tok@sh\endcsname{\def\PYG@tc##1{\textcolor[rgb]{0.73,0.13,0.13}{##1}}}
\expandafter\def\csname PYG@tok@ow\endcsname{\let\PYG@bf=\textbf\def\PYG@tc##1{\textcolor[rgb]{0.67,0.13,1.00}{##1}}}
\expandafter\def\csname PYG@tok@sx\endcsname{\def\PYG@tc##1{\textcolor[rgb]{0.00,0.50,0.00}{##1}}}
\expandafter\def\csname PYG@tok@bp\endcsname{\def\PYG@tc##1{\textcolor[rgb]{0.00,0.50,0.00}{##1}}}
\expandafter\def\csname PYG@tok@c1\endcsname{\let\PYG@it=\textit\def\PYG@tc##1{\textcolor[rgb]{0.25,0.50,0.50}{##1}}}
\expandafter\def\csname PYG@tok@fm\endcsname{\def\PYG@tc##1{\textcolor[rgb]{0.00,0.00,1.00}{##1}}}
\expandafter\def\csname PYG@tok@o\endcsname{\def\PYG@tc##1{\textcolor[rgb]{0.40,0.40,0.40}{##1}}}
\expandafter\def\csname PYG@tok@kc\endcsname{\let\PYG@bf=\textbf\def\PYG@tc##1{\textcolor[rgb]{0.00,0.50,0.00}{##1}}}
\expandafter\def\csname PYG@tok@c\endcsname{\let\PYG@it=\textit\def\PYG@tc##1{\textcolor[rgb]{0.25,0.50,0.50}{##1}}}
\expandafter\def\csname PYG@tok@mf\endcsname{\def\PYG@tc##1{\textcolor[rgb]{0.40,0.40,0.40}{##1}}}
\expandafter\def\csname PYG@tok@err\endcsname{\def\PYG@bc##1{\setlength{\fboxsep}{0pt}\fcolorbox[rgb]{1.00,0.00,0.00}{1,1,1}{\strut ##1}}}
\expandafter\def\csname PYG@tok@mb\endcsname{\def\PYG@tc##1{\textcolor[rgb]{0.40,0.40,0.40}{##1}}}
\expandafter\def\csname PYG@tok@ss\endcsname{\def\PYG@tc##1{\textcolor[rgb]{0.10,0.09,0.49}{##1}}}
\expandafter\def\csname PYG@tok@sr\endcsname{\def\PYG@tc##1{\textcolor[rgb]{0.73,0.40,0.53}{##1}}}
\expandafter\def\csname PYG@tok@mo\endcsname{\def\PYG@tc##1{\textcolor[rgb]{0.40,0.40,0.40}{##1}}}
\expandafter\def\csname PYG@tok@kd\endcsname{\let\PYG@bf=\textbf\def\PYG@tc##1{\textcolor[rgb]{0.00,0.50,0.00}{##1}}}
\expandafter\def\csname PYG@tok@mi\endcsname{\def\PYG@tc##1{\textcolor[rgb]{0.40,0.40,0.40}{##1}}}
\expandafter\def\csname PYG@tok@kn\endcsname{\let\PYG@bf=\textbf\def\PYG@tc##1{\textcolor[rgb]{0.00,0.50,0.00}{##1}}}
\expandafter\def\csname PYG@tok@cpf\endcsname{\let\PYG@it=\textit\def\PYG@tc##1{\textcolor[rgb]{0.25,0.50,0.50}{##1}}}
\expandafter\def\csname PYG@tok@kr\endcsname{\let\PYG@bf=\textbf\def\PYG@tc##1{\textcolor[rgb]{0.00,0.50,0.00}{##1}}}
\expandafter\def\csname PYG@tok@s\endcsname{\def\PYG@tc##1{\textcolor[rgb]{0.73,0.13,0.13}{##1}}}
\expandafter\def\csname PYG@tok@kp\endcsname{\def\PYG@tc##1{\textcolor[rgb]{0.00,0.50,0.00}{##1}}}
\expandafter\def\csname PYG@tok@w\endcsname{\def\PYG@tc##1{\textcolor[rgb]{0.73,0.73,0.73}{##1}}}
\expandafter\def\csname PYG@tok@kt\endcsname{\def\PYG@tc##1{\textcolor[rgb]{0.69,0.00,0.25}{##1}}}
\expandafter\def\csname PYG@tok@sc\endcsname{\def\PYG@tc##1{\textcolor[rgb]{0.73,0.13,0.13}{##1}}}
\expandafter\def\csname PYG@tok@sb\endcsname{\def\PYG@tc##1{\textcolor[rgb]{0.73,0.13,0.13}{##1}}}
\expandafter\def\csname PYG@tok@sa\endcsname{\def\PYG@tc##1{\textcolor[rgb]{0.73,0.13,0.13}{##1}}}
\expandafter\def\csname PYG@tok@k\endcsname{\let\PYG@bf=\textbf\def\PYG@tc##1{\textcolor[rgb]{0.00,0.50,0.00}{##1}}}
\expandafter\def\csname PYG@tok@se\endcsname{\let\PYG@bf=\textbf\def\PYG@tc##1{\textcolor[rgb]{0.73,0.40,0.13}{##1}}}
\expandafter\def\csname PYG@tok@sd\endcsname{\let\PYG@it=\textit\def\PYG@tc##1{\textcolor[rgb]{0.73,0.13,0.13}{##1}}}

\def\PYGZus{\char`\_}

\def\PYGZsh{\char`\#}

\def\PYGZhy{\char`\-}
\def\PYGZsq{\char`\'}

\makeatother

\makeatletter
\def\PYGdefault@reset{\let\PYGdefault@it=\relax \let\PYGdefault@bf=\relax%
    \let\PYGdefault@ul=\relax \let\PYGdefault@tc=\relax%
    \let\PYGdefault@bc=\relax \let\PYGdefault@ff=\relax}
\def\PYGdefault@tok#1{\csname PYGdefault@tok@#1\endcsname}
\def\PYGdefault@toks#1+{\ifx\relax#1\empty\else%
    \PYGdefault@tok{#1}\expandafter\PYGdefault@toks\fi}
\def\PYGdefault@do#1{\PYGdefault@bc{\PYGdefault@tc{\PYGdefault@ul{%
    \PYGdefault@it{\PYGdefault@bf{\PYGdefault@ff{#1}}}}}}}
\def\PYGdefault#1#2{\PYGdefault@reset\PYGdefault@toks#1+\relax+\PYGdefault@do{#2}}

\expandafter\def\csname PYGdefault@tok@gd\endcsname{\def\PYGdefault@tc##1{\textcolor[rgb]{0.63,0.00,0.00}{##1}}}
\expandafter\def\csname PYGdefault@tok@gu\endcsname{\let\PYGdefault@bf=\textbf\def\PYGdefault@tc##1{\textcolor[rgb]{0.50,0.00,0.50}{##1}}}
\expandafter\def\csname PYGdefault@tok@gt\endcsname{\def\PYGdefault@tc##1{\textcolor[rgb]{0.00,0.27,0.87}{##1}}}
\expandafter\def\csname PYGdefault@tok@gs\endcsname{\let\PYGdefault@bf=\textbf}
\expandafter\def\csname PYGdefault@tok@gr\endcsname{\def\PYGdefault@tc##1{\textcolor[rgb]{1.00,0.00,0.00}{##1}}}
\expandafter\def\csname PYGdefault@tok@cm\endcsname{\let\PYGdefault@it=\textit\def\PYGdefault@tc##1{\textcolor[rgb]{0.25,0.50,0.50}{##1}}}
\expandafter\def\csname PYGdefault@tok@vg\endcsname{\def\PYGdefault@tc##1{\textcolor[rgb]{0.10,0.09,0.49}{##1}}}
\expandafter\def\csname PYGdefault@tok@vi\endcsname{\def\PYGdefault@tc##1{\textcolor[rgb]{0.10,0.09,0.49}{##1}}}
\expandafter\def\csname PYGdefault@tok@vm\endcsname{\def\PYGdefault@tc##1{\textcolor[rgb]{0.10,0.09,0.49}{##1}}}
\expandafter\def\csname PYGdefault@tok@mh\endcsname{\def\PYGdefault@tc##1{\textcolor[rgb]{0.40,0.40,0.40}{##1}}}
\expandafter\def\csname PYGdefault@tok@cs\endcsname{\let\PYGdefault@it=\textit\def\PYGdefault@tc##1{\textcolor[rgb]{0.25,0.50,0.50}{##1}}}
\expandafter\def\csname PYGdefault@tok@ge\endcsname{\let\PYGdefault@it=\textit}
\expandafter\def\csname PYGdefault@tok@vc\endcsname{\def\PYGdefault@tc##1{\textcolor[rgb]{0.10,0.09,0.49}{##1}}}
\expandafter\def\csname PYGdefault@tok@il\endcsname{\def\PYGdefault@tc##1{\textcolor[rgb]{0.40,0.40,0.40}{##1}}}
\expandafter\def\csname PYGdefault@tok@go\endcsname{\def\PYGdefault@tc##1{\textcolor[rgb]{0.53,0.53,0.53}{##1}}}
\expandafter\def\csname PYGdefault@tok@cp\endcsname{\def\PYGdefault@tc##1{\textcolor[rgb]{0.74,0.48,0.00}{##1}}}
\expandafter\def\csname PYGdefault@tok@gi\endcsname{\def\PYGdefault@tc##1{\textcolor[rgb]{0.00,0.63,0.00}{##1}}}
\expandafter\def\csname PYGdefault@tok@gh\endcsname{\let\PYGdefault@bf=\textbf\def\PYGdefault@tc##1{\textcolor[rgb]{0.00,0.00,0.50}{##1}}}
\expandafter\def\csname PYGdefault@tok@ni\endcsname{\let\PYGdefault@bf=\textbf\def\PYGdefault@tc##1{\textcolor[rgb]{0.60,0.60,0.60}{##1}}}
\expandafter\def\csname PYGdefault@tok@nl\endcsname{\def\PYGdefault@tc##1{\textcolor[rgb]{0.63,0.63,0.00}{##1}}}
\expandafter\def\csname PYGdefault@tok@nn\endcsname{\let\PYGdefault@bf=\textbf\def\PYGdefault@tc##1{\textcolor[rgb]{0.00,0.00,1.00}{##1}}}
\expandafter\def\csname PYGdefault@tok@no\endcsname{\def\PYGdefault@tc##1{\textcolor[rgb]{0.53,0.00,0.00}{##1}}}
\expandafter\def\csname PYGdefault@tok@na\endcsname{\def\PYGdefault@tc##1{\textcolor[rgb]{0.49,0.56,0.16}{##1}}}
\expandafter\def\csname PYGdefault@tok@nb\endcsname{\def\PYGdefault@tc##1{\textcolor[rgb]{0.00,0.50,0.00}{##1}}}
\expandafter\def\csname PYGdefault@tok@nc\endcsname{\let\PYGdefault@bf=\textbf\def\PYGdefault@tc##1{\textcolor[rgb]{0.00,0.00,1.00}{##1}}}
\expandafter\def\csname PYGdefault@tok@nd\endcsname{\def\PYGdefault@tc##1{\textcolor[rgb]{0.67,0.13,1.00}{##1}}}
\expandafter\def\csname PYGdefault@tok@ne\endcsname{\let\PYGdefault@bf=\textbf\def\PYGdefault@tc##1{\textcolor[rgb]{0.82,0.25,0.23}{##1}}}
\expandafter\def\csname PYGdefault@tok@nf\endcsname{\def\PYGdefault@tc##1{\textcolor[rgb]{0.00,0.00,1.00}{##1}}}
\expandafter\def\csname PYGdefault@tok@si\endcsname{\let\PYGdefault@bf=\textbf\def\PYGdefault@tc##1{\textcolor[rgb]{0.73,0.40,0.53}{##1}}}
\expandafter\def\csname PYGdefault@tok@s2\endcsname{\def\PYGdefault@tc##1{\textcolor[rgb]{0.73,0.13,0.13}{##1}}}
\expandafter\def\csname PYGdefault@tok@nt\endcsname{\let\PYGdefault@bf=\textbf\def\PYGdefault@tc##1{\textcolor[rgb]{0.00,0.50,0.00}{##1}}}
\expandafter\def\csname PYGdefault@tok@nv\endcsname{\def\PYGdefault@tc##1{\textcolor[rgb]{0.10,0.09,0.49}{##1}}}
\expandafter\def\csname PYGdefault@tok@s1\endcsname{\def\PYGdefault@tc##1{\textcolor[rgb]{0.73,0.13,0.13}{##1}}}
\expandafter\def\csname PYGdefault@tok@dl\endcsname{\def\PYGdefault@tc##1{\textcolor[rgb]{0.73,0.13,0.13}{##1}}}
\expandafter\def\csname PYGdefault@tok@ch\endcsname{\let\PYGdefault@it=\textit\def\PYGdefault@tc##1{\textcolor[rgb]{0.25,0.50,0.50}{##1}}}
\expandafter\def\csname PYGdefault@tok@m\endcsname{\def\PYGdefault@tc##1{\textcolor[rgb]{0.40,0.40,0.40}{##1}}}
\expandafter\def\csname PYGdefault@tok@gp\endcsname{\let\PYGdefault@bf=\textbf\def\PYGdefault@tc##1{\textcolor[rgb]{0.00,0.00,0.50}{##1}}}
\expandafter\def\csname PYGdefault@tok@sh\endcsname{\def\PYGdefault@tc##1{\textcolor[rgb]{0.73,0.13,0.13}{##1}}}
\expandafter\def\csname PYGdefault@tok@ow\endcsname{\let\PYGdefault@bf=\textbf\def\PYGdefault@tc##1{\textcolor[rgb]{0.67,0.13,1.00}{##1}}}
\expandafter\def\csname PYGdefault@tok@sx\endcsname{\def\PYGdefault@tc##1{\textcolor[rgb]{0.00,0.50,0.00}{##1}}}
\expandafter\def\csname PYGdefault@tok@bp\endcsname{\def\PYGdefault@tc##1{\textcolor[rgb]{0.00,0.50,0.00}{##1}}}
\expandafter\def\csname PYGdefault@tok@c1\endcsname{\let\PYGdefault@it=\textit\def\PYGdefault@tc##1{\textcolor[rgb]{0.25,0.50,0.50}{##1}}}
\expandafter\def\csname PYGdefault@tok@fm\endcsname{\def\PYGdefault@tc##1{\textcolor[rgb]{0.00,0.00,1.00}{##1}}}
\expandafter\def\csname PYGdefault@tok@o\endcsname{\def\PYGdefault@tc##1{\textcolor[rgb]{0.40,0.40,0.40}{##1}}}
\expandafter\def\csname PYGdefault@tok@kc\endcsname{\let\PYGdefault@bf=\textbf\def\PYGdefault@tc##1{\textcolor[rgb]{0.00,0.50,0.00}{##1}}}
\expandafter\def\csname PYGdefault@tok@c\endcsname{\let\PYGdefault@it=\textit\def\PYGdefault@tc##1{\textcolor[rgb]{0.25,0.50,0.50}{##1}}}
\expandafter\def\csname PYGdefault@tok@mf\endcsname{\def\PYGdefault@tc##1{\textcolor[rgb]{0.40,0.40,0.40}{##1}}}
\expandafter\def\csname PYGdefault@tok@err\endcsname{\def\PYGdefault@bc##1{\setlength{\fboxsep}{0pt}\fcolorbox[rgb]{1.00,0.00,0.00}{1,1,1}{\strut ##1}}}
\expandafter\def\csname PYGdefault@tok@mb\endcsname{\def\PYGdefault@tc##1{\textcolor[rgb]{0.40,0.40,0.40}{##1}}}
\expandafter\def\csname PYGdefault@tok@ss\endcsname{\def\PYGdefault@tc##1{\textcolor[rgb]{0.10,0.09,0.49}{##1}}}
\expandafter\def\csname PYGdefault@tok@sr\endcsname{\def\PYGdefault@tc##1{\textcolor[rgb]{0.73,0.40,0.53}{##1}}}
\expandafter\def\csname PYGdefault@tok@mo\endcsname{\def\PYGdefault@tc##1{\textcolor[rgb]{0.40,0.40,0.40}{##1}}}
\expandafter\def\csname PYGdefault@tok@kd\endcsname{\let\PYGdefault@bf=\textbf\def\PYGdefault@tc##1{\textcolor[rgb]{0.00,0.50,0.00}{##1}}}
\expandafter\def\csname PYGdefault@tok@mi\endcsname{\def\PYGdefault@tc##1{\textcolor[rgb]{0.40,0.40,0.40}{##1}}}
\expandafter\def\csname PYGdefault@tok@kn\endcsname{\let\PYGdefault@bf=\textbf\def\PYGdefault@tc##1{\textcolor[rgb]{0.00,0.50,0.00}{##1}}}
\expandafter\def\csname PYGdefault@tok@cpf\endcsname{\let\PYGdefault@it=\textit\def\PYGdefault@tc##1{\textcolor[rgb]{0.25,0.50,0.50}{##1}}}
\expandafter\def\csname PYGdefault@tok@kr\endcsname{\let\PYGdefault@bf=\textbf\def\PYGdefault@tc##1{\textcolor[rgb]{0.00,0.50,0.00}{##1}}}
\expandafter\def\csname PYGdefault@tok@s\endcsname{\def\PYGdefault@tc##1{\textcolor[rgb]{0.73,0.13,0.13}{##1}}}
\expandafter\def\csname PYGdefault@tok@kp\endcsname{\def\PYGdefault@tc##1{\textcolor[rgb]{0.00,0.50,0.00}{##1}}}
\expandafter\def\csname PYGdefault@tok@w\endcsname{\def\PYGdefault@tc##1{\textcolor[rgb]{0.73,0.73,0.73}{##1}}}
\expandafter\def\csname PYGdefault@tok@kt\endcsname{\def\PYGdefault@tc##1{\textcolor[rgb]{0.69,0.00,0.25}{##1}}}
\expandafter\def\csname PYGdefault@tok@sc\endcsname{\def\PYGdefault@tc##1{\textcolor[rgb]{0.73,0.13,0.13}{##1}}}
\expandafter\def\csname PYGdefault@tok@sb\endcsname{\def\PYGdefault@tc##1{\textcolor[rgb]{0.73,0.13,0.13}{##1}}}
\expandafter\def\csname PYGdefault@tok@sa\endcsname{\def\PYGdefault@tc##1{\textcolor[rgb]{0.73,0.13,0.13}{##1}}}
\expandafter\def\csname PYGdefault@tok@k\endcsname{\let\PYGdefault@bf=\textbf\def\PYGdefault@tc##1{\textcolor[rgb]{0.00,0.50,0.00}{##1}}}
\expandafter\def\csname PYGdefault@tok@se\endcsname{\let\PYGdefault@bf=\textbf\def\PYGdefault@tc##1{\textcolor[rgb]{0.73,0.40,0.13}{##1}}}
\expandafter\def\csname PYGdefault@tok@sd\endcsname{\let\PYGdefault@it=\textit\def\PYGdefault@tc##1{\textcolor[rgb]{0.73,0.13,0.13}{##1}}}

\makeatother

\usepackage{natbib}

\usepackage{framed}

\usepackage{algorithm}
\usepackage{algorithmic}

\usepackage{inconsolata}
\usepackage{hyperref}

\usepackage[accepted]{icml2017}
\icmltitlerunning{Kapre: Keras Audio Preprocessing Layers}

\begin{document} 

\twocolumn[
\icmltitle{Kapre: On-GPU Audio Preprocessing Layers for a Quick Implementation of Deep Neural Network Models with Keras}

\begin{icmlauthorlist}
\icmlauthor{Keunwoo Choi}{qmul}
\icmlauthor{Deokjin Joo}{il}
\icmlauthor{Juho Kim}{il}
\end{icmlauthorlist}

\icmlaffiliation{qmul}{Centre for Digital Music, Queen Mary University of London, London, UK}
\icmlaffiliation{il}{University of Illinois at Urbana-Champaign, USA}

\icmlcorrespondingauthor{Keunwoo Choi}{keunwoo.choi@qmul.ac.uk}

\icmlkeywords{audio preprocessing, deep learning, keras, kapre, music}

\vskip 0.3in
]

\printAffiliationsAndNotice{}  
% ABSTRACT
\begin{abstract}
We introduce \textit{Kapre}, Keras layers for audio and music signal preprocessing. Music research using deep neural networks requires a heavy and tedious preprocessing stage, for which audio processing parameters are often ignored in parameter optimisation. To solve this problem, Kapre implements time-frequency conversions, normalisation, and data augmentation as Keras layers. We report simple benchmark results, showing real-time on-GPU preprocessing adds a reasonable amount of computation.

\end{abstract}

% SECTION: INTRODUCTION
\section{Introduction}
Deep learning approach has been gaining attention in music and audio informatics research, achieving state-of-the-art performances in many problems including audio event detection \cite{aytar2016soundnet} and music tagging \cite{choi2016automatic}. 

Since building deep neural network models is becoming easier using frameworks that provide off-the-shelf modules, e.g., Keras \citep{chollet2015}, preprocessing data often occupies lots of time and effort. It is more problematic when dealing with audio data than images or texts due to its large size and heavy decoding computation. A procedure for audio data preparation generally includes \textit{i}) decoding, \textit{ii}) resampling, and \textit{iii}) conversion to a time-frequency representation. Stages \textit{i}/\textit{ii} should be readily done since otherwise they would be a huge bottleneck. However, Stage \textit{iii} can be implemented in many ways with pros and cons.

Running Stage \textit{iii} in whether real-time or not is a trade-off between storage and computation time. \textbf{Pros}: It enables to search the best audio preprocessing configuration e.g., time-frequency representations and their parameters. It can vastly save storage, which usually is needed as much as decoded audio samples for each configuration. \textbf{Cons}: Additional computation may impede training and inference.

One of the main reasons to propose an on-GPU audio preprocessing is for a quick and easy implementation. Adding a preprocessing layer can be done by a single line of code. It can be done and may be faster on CPU with multiprocessing, but an optimised implementation is challenging.

With {\em Kapre}\footnote{\url{https://github.com/keunwoochoi/kapre}}, the whole preparation and training procedure becomes simple: \textit{i}) Decode (and possibly resample) audio files and save them as binary formats, \textit{ii}) implement a generator that loads the data, and \textit{iii}) add a Kapre layer at the input side of Keras model. In this way, the researcher's \textit{user experience} is improved as well as audio processing parameters can be optimised.

A relevant question on the proposed practice is how much time this approach would take for those merits. We present the results of a simple benchmark in Section~\ref{sec:exp}.

% CODE
\begin{listing}[t]
\begin{framed}\footnotesize
\vspace{-2mm}
%\begin{minted}[fontsize=\footnotesize]{python} 
%model = Sequential()
%model.add(Melspectrogram(
%    input_shape=(2, 44100), # 1-sec stereo input
%    n_dft=512, n_hop=256, n_mels=128, sr=sr,
%    fmin=0.0, fmax=sr/2, return_decibel=False,
%    trainable_fb=False, trainable_kernel=False))
%model.add(Normalization2D(str_axis='freq'))
%model.add(AdditiveNoise(power=0.2))
%# and more layers for model hereafter
%\end{minted}
\begin{Verbatim}[commandchars=\\\{\}]
\PYG{n}{model} \PYG{o}{=} \PYG{n}{Sequential}\PYG{p}{()}
\PYG{n}{model}\PYG{o}{.}\PYG{n}{add}\PYG{p}{(}\PYG{n}{Melspectrogram}\PYG{p}{(}
    \PYG{n}{input\PYGZus{}shape}\PYG{o}{=}\PYG{p}{(}\PYG{l+m+mi}{2}\PYG{p}{,} \PYG{l+m+mi}{44100}\PYG{p}{),} \PYG{c+c1}{\PYGZsh{} 1\PYGZhy{}sec stereo input}
    \PYG{n}{n\PYGZus{}dft}\PYG{o}{=}\PYG{l+m+mi}{512}\PYG{p}{,} \PYG{n}{n\PYGZus{}hop}\PYG{o}{=}\PYG{l+m+mi}{256}\PYG{p}{,} \PYG{n}{n\PYGZus{}mels}\PYG{o}{=}\PYG{l+m+mi}{128}\PYG{p}{,} \PYG{n}{sr}\PYG{o}{=}\PYG{n}{sr}\PYG{p}{,}
    \PYG{n}{fmin}\PYG{o}{=}\PYG{l+m+mf}{0.0}\PYG{p}{,} \PYG{n}{fmax}\PYG{o}{=}\PYG{n}{sr}\PYG{o}{/}\PYG{l+m+mi}{2}\PYG{p}{,} \PYG{n}{return\PYGZus{}decibel}\PYG{o}{=}\PYG{n+nb+bp}{False}\PYG{p}{,}
    \PYG{n}{trainable\PYGZus{}fb}\PYG{o}{=}\PYG{n+nb+bp}{False}\PYG{p}{,} \PYG{n}{trainable\PYGZus{}kernel}\PYG{o}{=}\PYG{n+nb+bp}{False}\PYG{p}{))}
\PYG{n}{model}\PYG{o}{.}\PYG{n}{add}\PYG{p}{(}\PYG{n}{Normalization2D}\PYG{p}{(}\PYG{n}{str\PYGZus{}axis}\PYG{o}{=}\PYG{l+s+s1}{\PYGZsq{}freq\PYGZsq{}}\PYG{p}{))}
\PYG{n}{model}\PYG{o}{.}\PYG{n}{add}\PYG{p}{(}\PYG{n}{AdditiveNoise}\PYG{p}{(}\PYG{n}{power}\PYG{o}{=}\PYG{l+m+mf}{0.2}\PYG{p}{))}
\PYG{c+c1}{\PYGZsh{} and more layers for model hereafter}
\end{Verbatim}

\vspace{-5mm}
\end{framed}
\caption{A code snippet that computes Mel-spectrogram, normalises per frequency, and adds Gaussian noise.}
\label{code}
\vspace{-0.3cm}
\end{listing}

Librosa \cite{mcfee_brian_2017_293021} and Essentia \cite{bogdanov2013essentia} provide audio and music analysis on Python. They can be combined with input preprocessing utility such as Pescador \footnotemark\footnotetext{\url{http://pescador.readthedocs.io/}} and Fuel \cite{van2015blocks} to implement an effective data pipeline. PyTorch has its own audio file loader\footnote{\url{https://github.com/pytorch/audio}} but it does not support audio analysis such as computing spectrograms.

% SECTION: KAPRE
\section{Kapre}
As mentioned earlier, the main goal of Kapre is to perform audio preprocessing in Keras layers. Listing~\ref{code} is an example code that computes Mel-spectrogram, normalises it per frequency, and adds Gaussian noise. Several important layers are summarised as below and available as of Kapre version 0.1.

$\bullet$\hspace{1.5mm}\texttt{Spectrogram} uses two 1-dimensional convolutions, each of which is initialised with real and imaginary part of discrete Fourier transform kernels respectively, following the definition of discrete Fourier transform: 
\begin{equation}
X_k=\sum_{n=0}^{N-1}x_n\cdot[\cos(2\pi kn/N)-i\cdot \sin(2\pi kn/N)] 
\end{equation}
for $k\in~[0,N-1]$. 
The computation is implemented with \texttt{conv2d} of Keras backend, which means the Fourier transform kernels can be trained with backpropagation. 

$\bullet$\hspace{1.5mm}\texttt{Melspectrogram} is an extended layer based on \texttt{Spectrogram} with a multiplication by mel-scale conversion matrix from linear frequencies which can be trained.

$\bullet$\hspace{1.5mm}\texttt{Normalization2D} normalises 2D input data (e.g., time-frequency representations) per frequency, time, channel, (single) data, and batch.

$\bullet$\hspace{1.5mm}\texttt{Filterbank} provides a general filterbank layer that can be initialised with mel/log/linear frequency scales as well as random.

$\bullet$\hspace{1.5mm}\texttt{AdditiveNoise} adds different types of noise for data augmentation. The noise gain can be randomised for further augmentation. Noise is only applied in training phase.

% SECTION: EXPERIMENTS
\section{Experiments and Conclusions}\label{sec:exp}
\begin{table}[tbp]
\vspace{-0.4cm}
\centering
\caption{A 5-layer convolutional neural network used in the experiment.}
\label{table}
\begin{tabular}{c|l|l}
\begin{tabular}[c]{@{}c@{}}Layer index\end{tabular} & Layer type              & Note                                                                          \\ \hline
1                                                     & Conv2D(64, (20, 3)) & \begin{tabular}[c]{@{}l@{}}ReLU activation\\ (2, 2) stride\end{tabular}       \\ \hline
2 -- 5                                                   & Conv2D(64, (3, 3))  & \begin{tabular}[c]{@{}l@{}}ReLU activation\\ (2, 2) stride\end{tabular}       \\ \hline
6                                                     & AveragePooling2D()  & Global average                                                                              \\ \hline
7                                                     & Dense               & \begin{tabular}[c]{@{}l@{}}88 output nodes,\\ Softmax activation\end{tabular}
\end{tabular}
\vspace{-0.3cm}
\end{table}

\begin{figure}
\includegraphics[width=\columnwidth]{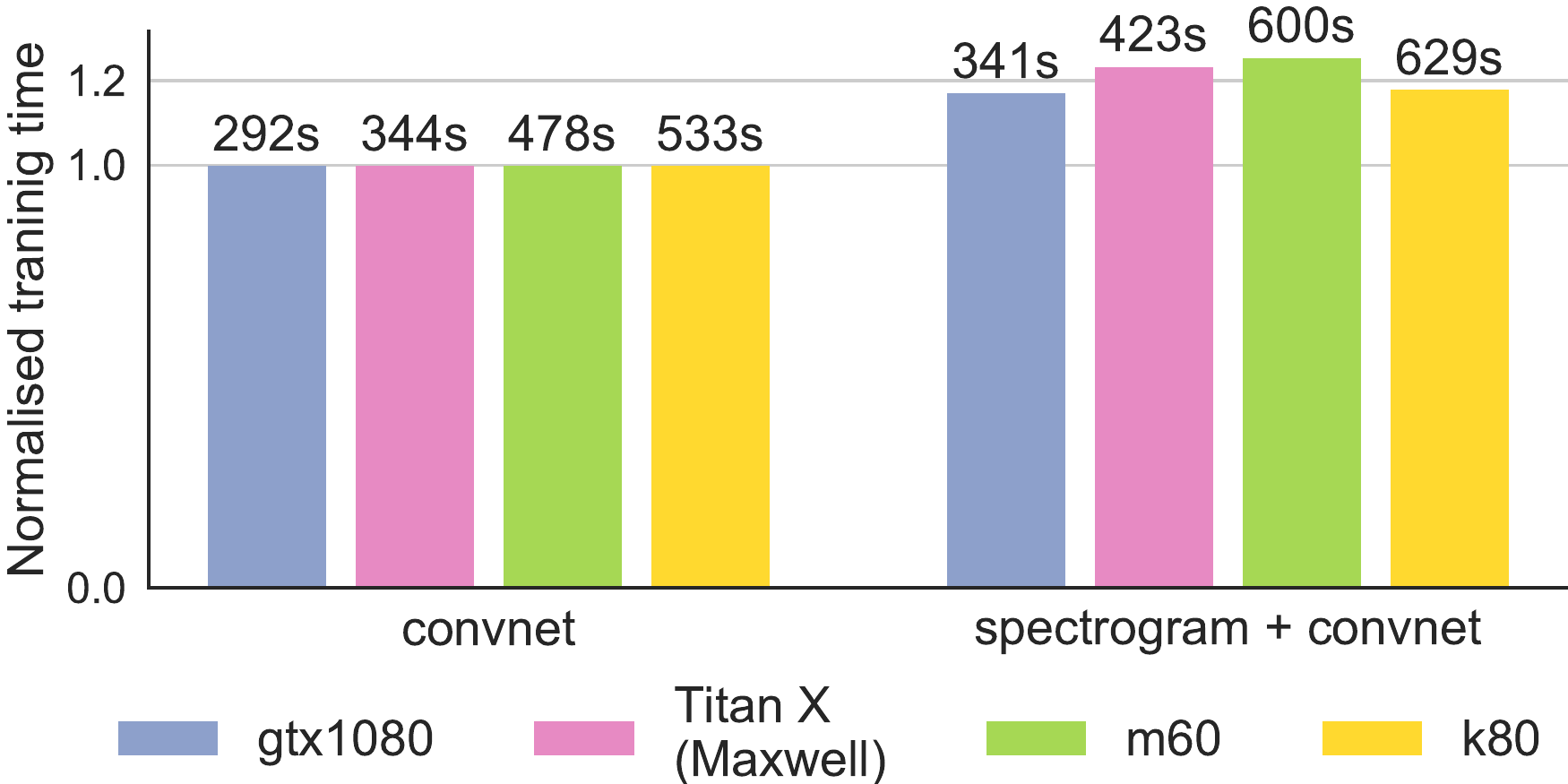}
\caption{Normalised time consumed to train a convnet with and without audio preprocessing layer}\label{fig:result}
\end{figure}

% \textbf{Goal of experiments}
An experiment is designed to investigate the additional computation that Kapre introduces during training. We do not directly measure the computation time of Kapre preprocessing. Instead, we measure the time difference between `time-frequency conversion + convnet' vs. `convnet' because these are realistic scenarios if we consider pre-computing spectrograms as an alternative. This comparison compensates the effects of potential overheads e.g., i/o.

A 5-layer convolutional neural network summarised in Table~\ref{table} is used. We use a dummy input/output data that is generated before training and simulates 30-second mono signal with 32,000 Hz sampling rate/88D one-hot-vector respectively. There are 157,336 parameters in this network, which is relatively small compared to many models recently used in music/audio research. The training is configured as batch size of 16, 512 batches per epoch, and 2 epochs, iterating 16,384 training samples overall. The experiment is implemented with Keras 2.0.4 \cite{chollet2015}, Theano 0.9.0 \cite{2016arXiv160502688short}, CUDA 8, and cuDNN 7.

For time-frequency conversion, short-time Fourier transform with 512-point DFT, 50\% overlap, and decibel scaling is computed using \texttt{kapre.time\_frequency.Spectrogram} layer.

% \textbf{results}
Figure \ref{fig:result} compares the time consumptions normalised by without-conversion cases and on different GPUs. In the experiment, the audio preprocessing layer adds about 20\% training time. Changing batch size did not affect this proportion.  Although it is plotted after normalisation for devices, we should focus on the absolute time difference, too, because that is the time consumed regardless of the size of deep neural network model. This means on-GPU preprocessing is more suitable for large-scale work since the additional time cost, which is constant, can be relatively minor for the training of bigger networks while benefiting on storage usage with potentially larger-scale dataset.

To conclude, we proposed to adopt on-GPU audio preprocessing for faster and easier prototyping. Kapre layers enables a storage-efficient input preprocessing optimisation. We presented the additional computation time due to the preprocessing on GPU which can be relatively small when large networks are used. In practice, data may readily be preprocessed for more efficient model hyperparameter search once preprocessing parameters are set, until which the proposed method can be efficient. 

\bibliography{kapre}

\bibliographystyle{icml2017}

\end{document}